\documentclass[prl,twocolumn]{revtex4}
\usepackage{psfig}
\def \frac#1#2{ { #1 \over #2} }
\begin{document}

\title{Many-body effects in nuclear structure}


\author{F. Barranco $^{1}$, 
R.A. Broglia $^{2,3,4}$, G. Col\`o $^{2,3}$,
E. Vigezzi $^{3}$, P.F. Bortignon$^{2,3}$}

\address{ $^1$ Departamento de Fisica Aplicada III, Universidad de Sevilla,
Spain}

\address{$^2$ Dipartimento di Fisica, Universit\`a di  Milano, Italy}

\address{$^3$ INFN Sezione di Milano, Italy}

\address{$^4$ The Niels Bohr Insitute, University of Copenhagen, Denmark}

\begin{abstract}

We calculate, for the first time, 
the state-dependent pairing gap of a finite nucleus ($^{120}$Sn)
diagonalizing the bare  nucleon-nucleon potential (Argonne $v_{14}$)
in a Hartree-Fock
basis (with effective $k-$mass $m_k \approx$ 0.7 m), within the framework
of the BCS approximation including scattering states up to 800 MeV above
the Fermi energy to achieve convergence. The resulting gap accounts for about
half of the experimental gap. 
We find that
a consistent description of the low-energy nuclear spectrum requires, aside
from the bare nucleon-nucleon interaction,  not only
the dressing of single-particle motion through the coupling to the nuclear
surface, to give the right density of levels close to the Fermi energy
(and thus an effective mass $m^* \approx m$), 
but also the renormalization of collective vibrational modes
through vertex and self-energy processes, processes which are also found
to play an essential role in the pairing channel, leading to a long range, state dependent component 
of the pairing interaction. 
The combined effect of the bare nucleon-nucleon
potential and of the induced pairing interaction arising 
from the exchange of low-lying surface vibrations between nucleons 
moving in time reversal states close to the Fermi energy 
accounts for the experimental
gap.
\end{abstract}

\maketitle
In the study of finite many-body systems such as the atomic nucleus
with its rich variety of quantal size effects, structural properties, and
fluctuations, the central problem has been to identify the appropriate degrees
of freedom  for describing the phenomena encountered. The complementary concepts
referring to the independent motion of the individual nucleons and the
collective behaviour of the nucleus as a whole provide the elementary modes
of excitation needed to describe the system \cite{[1]}. The unifying picture emerging
from the interweaving of these degrees of freedom
and described in terms of nuclear field theory (NFT) [2-6] 
based on the particle-vibration coupling
(for other related
 approaches, see e.g. [7-10]) and
tailored upon QED [11-12],
 has been applied to a number of schematic models and realistic situations
 [13-21] and its validity demonstrated. It thus provides a natural
 framework to assess the role different degrees of freedom play in
 the nuclear structure.
 
 An important subject presently under intensive study concerns the characterization 
 of an eventual long range component of the pairing interaction in nuclei
 [22-24]. In what follows we use NFT to assess the importance the exchange
 of vibrations between pairs of nucleons moving in time reversal states have in
 building up pairing correlations in nuclei, taking also into account
 self-energy and vertex corrections (i.e. avoiding using approximations
 which, in condensed matter literature are connected with the so-called 
 Migdal theorem, cf. ref. [25] and refs. therein).


To this scope, we study 
the quasiparticle and vibrational spectrum  of odd- and even-
isotopes of single-closed-shell nuclei, where all the richness  of the
single-particle and  collective degrees of freedom are fully expressed, avoiding the extra
complications of static deformations and associated rotations. The spectra
of the $^A_{50}$Sn isotopes, in particular those with mass number
$A$=119,120 and
121, with their abundance of detailed experimental information,
provide an excellent laboratory where to test the importance of the residual
pairing interaction and its relation to self-energy processes. To be
remembered that BCS theory connects the mass enhancement factor
$\lambda$ associated with the $\omega-$mass 
($m_{\omega}= m(1+ \lambda)$, cf, e.g. [26])   
with the pairing gap
$\Delta \approx 2 \omega_c exp(-1/\lambda)$, $\omega_c$
being the energy associated with a typical vibration (boson) of the system [25].

In general one fixes the parameters of the effective interaction of nucleons
in the nucleus, by requiring mean field theory, as a rule Hartree-Fock
or Hartree-Fock-Bogoliubov theory if the system is superfluid, to reproduce the
experimental findings: binding energies, mean square radii, etc.
This is equivalent to requiring that the solution of the
Schr\"odinger equation describing the
bound states of the electron-proton system, interacting through the Coulomb
force, reproduces the energy levels of the hydrogen atom.
We know that this is not possible, unless
the renormalization effects arising from the electron-photon
coupling are properly taken into account as prescribed by QED [11,12].
Similarly, the parameters of the effective nuclear interaction should reproduce
the experimental findings only when the particle-vibration coupling is allowed
to renormalize, screen and dress the different modes of elementary excitation
and the interaction among them in a similar way as, for example, 
the Lamb shift of the
hydrogen atom is
accounted for only when the renormalization effects arising from the
electron-photon coupling are considered.

The formalism we shall use is based on the Dyson equation
[22]. It can describe on
equal footing the dressed one-particle state $\tilde a $  of an
odd nucleon renormalized
by the (collective) response of all the other nucleons (Figs. 1(a)-1(d)), 
the renormalization of the energy $\hbar \omega_{\nu}$ (Figs. 2(a)-2(b))
and of the transition probability $B(E\lambda)$ (Figs. 2(c)-2(f))
of the collective vibrations of the even system 
where the number of nucleons remains constant (correlated
particle-hole excitations), and the induced interaction due to the exchange of
collective vibrations between pairs of nucleons [23], moving in time reversal
states close to the Fermi energy (Figs. 1(e)-1(g)).
We include both self-energy and vertex correction processes, thus 
satisying Ward identities (cf., e.g., [25]).
Within this framework,  the self-consistency existing
between the dynamical deformations of the density and of the potential
sustained by "screened" particle-vibrations coupling vertices leads to
renormalization effects which make  finite (stabilize)  the collectivity
and the self-interaction of the elementary modes of nuclear excitation,
in particular of the low-lying surface vibrational modes,
providing an accurate description of many seemingly unrelated experimental
findings, in terms of very few (theoretically calculable) parameters, namely:
the $k-$mass $m_k$ [26] and  the particle vibration coupling vertex 
$h(ab\nu)$, associated to the process in which a quasiparticle 
changes its state of motion from the
unperturbed quasiparticle state $a$ to $b$, by absorbing or emitting a 
vibration $\nu$ [1].

The equation describing the renormalization of a quasiparticle $a$,
due to this variety of couplings is

\begin{eqnarray}
\left [
\left( \matrix
{
E_a  &  0 \cr
0  &  -E_a \cr   } \right )
+
\left( \matrix
{
\Sigma_{11}(\tilde E_a)   &  \Sigma_{12}(\tilde E_a) \cr
 \Sigma_{12}(\tilde E_a)  &  \Sigma_{22}(\tilde E_a) \cr   } \right )
 \right ]
 \left ( \matrix {
\tilde x_a \cr
\tilde y_a \cr}
\right )  =  \nonumber \\
{\tilde E_a}
\left ( \matrix {
\tilde x_a \cr
\tilde y_a \cr}
\right ), \quad \quad \quad \quad \quad \quad \quad \quad  
\end{eqnarray}
where $\Sigma_{ii}$ and  
$\Sigma_{ij} , (i \neq j) $ are the normal and abnormal self-energies.


Eq. (1)is to
be solved iteratively, and simultaneously for all the involved quasiparticle
states. At each iteration step, the original 
quasiparticle states $a$ with occupation numbers $u_a$ and $v_a$, 
become fragmented over the different
eigenstates $\tilde a $ with probability $\tilde u_a^2 
+ \tilde v_a^2$,   while 
the renormalized occupation factors are obtained from 
the components of the eigenvectors, $\tilde x_a$ and $ \tilde y_a$, 
according to the relations
$\tilde u_a = \tilde x_a u_a + \tilde y_a v_a \;\;\;,\;\;\;\; 
\tilde v_a =  - \tilde y_a u_a + \tilde x_a v_a.$
The quantities $\tilde u_a$ and $\tilde v_a$ are related to the 
spectroscopic factors measured in one-nucleon stripping and pick-up  
reactions, respectively. One can also define [22,25] a 
renormalized state-dependent pairing gap,
through the relation $\tilde \Delta_{a} = 2\tilde E_a \tilde u_a \tilde v_a /(\tilde u_a^2
+ \tilde v_a^2)$, which in the limit of no fragmentation reduces to the usual
BCS expression [27].



\begin{figure}
\centerline{
\psfig{file=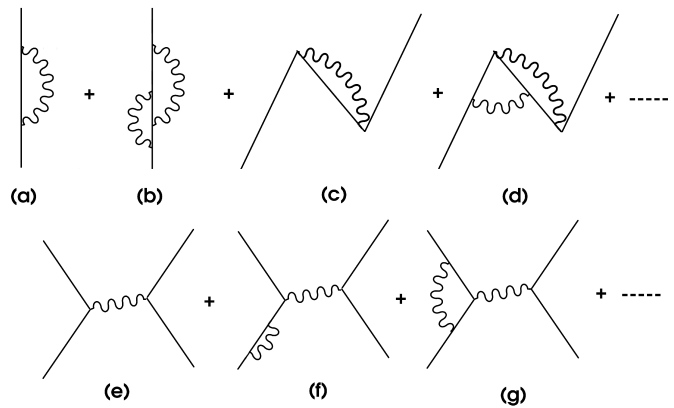,height=5.5cm,angle=0}}
\caption{
Renormalization processes arising from the particle vibration coupling
phenomenon. A line indicates quasiparticles obtained from BCS theory,
making use of the mean field single-particle states of Sly4
and the nucleon-nucleon $v_{14}$ 
Argonne potential. The wavy line indicates the vibrational states.}
\end{figure}

In the calculations reported below, a Skyrme interaction (Sly4 parametrization,
with $m_k \approx 0.7 m $[29]), was solely used to determine the properties of
the  bare single-particle states and the collective vibrations in the
particle-hole channel. Both the bare nucleon-nucleon $v_{14}$ Argonne 
potential as well as the exchange of collective vibrations were used in 
the particle-particle (pairing) channel.

As seen from Fig. 3,
Hartree-Fock theory is not able to account for the experimental 
quasiparticle energies of the low-lying states.
Diagonalizing the Argonne $v_{14}$ nucleon-nucleon potential in the
Hartree-Fock
basis, within the framework of the generalized Bogoliubov-Valatin approximation
including scattering states up to 800 MeV above the Fermi energy
(to achieve convergence)
in a spherical box of radius equal to 15 fm, one obtains the state-dependent 
pairing gap shown in Fig. 4 (labelled $v_{14})$.
The resulting pairing gap (average value for levels around the Fermi energy)
accounts for about half of the empirical
pairing gap value ($\approx$ 1.4 MeV) obtained from the odd-even mass difference 
[30]. 
In keeping with this result, the quasiparticle spectrum (cf. Fig. 3),
although being slightly closer to the experimental findings than that predicted by
Hartree-Fock theory, displays 
large discrepancies with observations.
The situation is rather similar concerning the low-lying
quadrupole vibration of $^{120}$Sn calculated in the QRPA
with standard effective nucleon-nucleon interactions like
Gogny or Skyrme forces. While energy is predicted too high,
which may not be too important,
the B(E2) value is about a factor 3 too
small (cf. Table 1), a result which calls for a better theory. 

In fact,
renormalizing the energy and the transition 
strength of the $2^+$
phonon, following NFT [2,3], that is, considering the couplings
of the type depicted in Fig. 2 (cf. also ref. [32] and refs. therein), 
one obtains 
an increase of the $B(E2)$ transition probability
which brings theory essentially in  agreement  with experiment (cf. Table 1)
[33].
The most important processes which renormalize the energy of the phonon 
are shown in Figs. 2(a) and (b). 
Other graphs which are also  of fourth order in the particle-vibration
coupling vertex, 
but contain intermediate states with more than four quasiparticle
states, lead to very small contributions. This is because these terms
 not only involve larger denominators, but also,
due to their higher degree of complexity, give rise  
to contributions with
"random" signs which tend to cancel each other. This is a consequence
of the fact that while cancellation between the contribution
associated wih graphs (a) and (b) of Fig. 2 is strong in the particle-hole
channel  the opposite is true in the particle-particle channel [35], and 
that the phonons are calculated in a Bogoliubov-Valatin-quasiparticle
basis. In keeping with the above discussion,
the most important processes renormalizing 
the B(E2) transition probability are those shown in Figs. 2(c),(d),(e) and (f).

\begin{figure}
\centerline{
\psfig{file=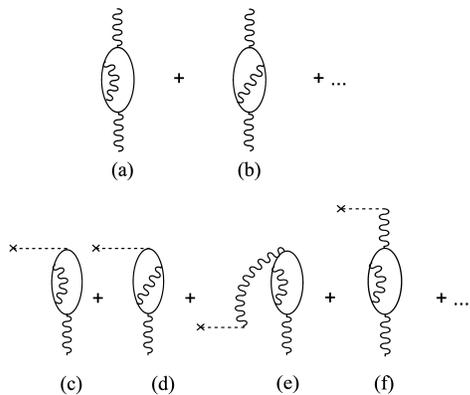,height=5.5cm}}
\caption{
Most relevant processes taken into account in the renormalization of the
energy of the phonon (a-b) and of the associated  transition strength (c-f). }
\end{figure}

\begin{figure}
\centerline{
\psfig{file=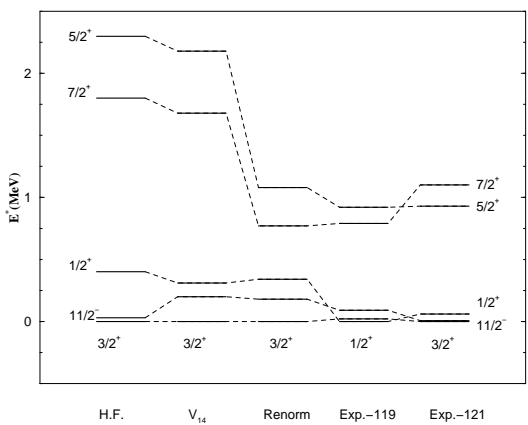,height=5.5cm,angle=-90}}
\caption{
The spectra of the lowest quasiparticle states in $^{120}$Sn  calculated
using Hartree-Fock theory, BCS with the Argonne $v_{14}$ potential,
and after renormalization, are 
compared to the experimental levels in the odd neighbouring nuclei $^{119}$Sn
and $^{121}$Sn.}
\end{figure}

We have also calculated 
the static quadrupole moment $Q$ of the $2^+$ state, including the 
contributions from the processes shown in Fig. 6.27 of ref. [1], considering
also self-energy effects [36].
The resulting value of $Q$ is rather small (8 e fm$^2$), 
in agreement  with the experimental findings  ($10 \pm 10$
e fm$^2$, or  --5 $\pm$ 10 e fm$^2$ [37,38] ).

Because in the above calculations we have included only a partial set 
(although the most important for the physics under discussion) of the NFT
graphs needed to provide a completely consistent description of
single-particle and collective vibration renormalizations, the mixing of spurious states 
with the physiscal states has to be contemplated. Although it is difficult
to give a precise estimate of the error induced by such undesired couplings, 30\%
effects have been found in the calculation of the energy of the one-phonon state [39].


Making use of phonons which account for the
experimental findings, the normal and abnormal 
self-energies were calculated, and Eq.(1) solved. The
average value of the resulting state-dependent pairing gap of $^{120}$Sn
is now close to the value $\Delta_{exp}= 1.4$ MeV 
derived from the odd-even mass difference
(cf. Fig. 4) [40]. In Fig. 3 we show the energy of the peaks 
carrying the largest quasiparticle strength, for the orbitals around the 
Fermi energy, which provide an
overall account of the lowest quasiparticle states measured
in the odd systems $^{119}$Sn and $^{121}$Sn. 

\begin{figure}
\centerline{
\psfig{file=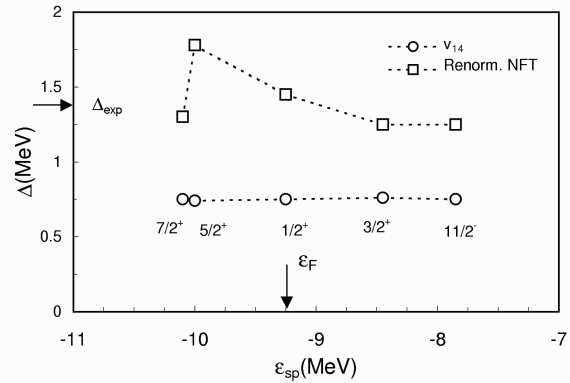,height=5.5cm,angle=0}}
\caption{
The state-dependent pairing gap for the levels close to the Fermi energy
obtained using BCS theory with the $v_{14}$ Argonne potential (circles)
is compared with the result obtained including renormalization effects
(squares).}
\end{figure}


One can conclude that
mean field theory and bare nucleon-nucleon potentials reproduce neither 
the experimental transition
strengths nor the  pairing gaps, 
let alone the density of quasiparticle states close to the ground state.
Dressing the single-particle motion, the correlated particle-hole excitations
of mean field and the nucleon-nucleon interaction with collective surface
vibrations, brings theory in overall agreeement with experiment.
In particular, about half of the pairing gap arises from the long range 
component of the pairing interaction associated with the exchange of
collective vibrations. To further clarify the interdependence of single-particle
and collective degrees of freedom, future studies should, for example, concentrate on the role
this interdependence has on the nuclear masses. In particular, whether 
the explicit, simplified, inclusion of ground state correlations and of
the induced pairing interaction can reduce the present r.m.s. error of 
0.674 MeV with which one of the best presently available Hartree-Fock mass formula [24]
is able to reproduce the experimental findings.    

\vskip 1cm






\vskip 1cm

\begin{table}[h]
\begin{tabular}{|c|c|c|}
\cline{2-3}
\multicolumn{1}{c|}{}  & \quad $\hbar \omega_{2+} $ (MeV)  \quad & B(E2 $\uparrow$) (e$^2$ fm$^4$)\\
\hline
 RPA (Gogny)& 2.9 & 660\\
\hline
 RPA  (Sly4) & 1.5 & 890\\
\hline
 RPA + renorm  &  0.9 &   2150\\
\hline
 Exp.   & 1.2 & 2030\\
\hline
\end{tabular}
\caption{The energy and reduced E2 transition strength of the low-lying
 2$^+$ state, calculated according to different theoretical models, are
 compared to the experimental values [37].
}
\end{table}









\begin{thebibliography}{99}
\bibitem{[27]} The formalism leading to Eq. (1)  
is equivalent to that 
presented in ref. [22], which was 
based on Green functions, and is closely connected to that of ref. [28],
based on the equation of motion method. 
To solve Eq.(1) we have diagonalized an
equivalent energy independent matrix as explained in ref. [28].
\bibitem{[30]} This result can be compared to  
a similar calculation
performed with  an effective mass equal to one, 
which gives $\Delta \approx$  2.2 MeV [31], a result which reflects the fact 
that  the gap
depends strongly on the density of levels at the Fermi energy.
\bibitem{[33]} We have only considered the renormalization of
the $2+$ low-lying phonon, the properties of the other phonons participating
in the renormalization processes being instead taken from experiment.
 For simplicity, the calculations  were carried out
making use of a separable multipole-multipole interaction, including the
modes with $\lambda^{\pi} =2^+,3^-,4^+,5^-$. The
coupling constant of the quadrupole-quadrupole component was 
adjusted so that the bare quadrupole  vibration displayed properties similar
to those calculated making use of effective interactions (i.e. $\hbar \omega_{2+} = 2$ MeV
and $B(E2 \uparrow) $ = 700 e$^2$ fm$^4$), while 
the other coupling constants were chosen, so as 
to give an overall description of the measured energies and 
transition strengths of the  low-lying states [34].
\bibitem{[36]} To be noted that the anharmonic effects associated to the finite
value of the quadrupole moment produce a shift of the 
energy of phonon, which is however very small compared to the processes
already considered.
\bibitem{[40]} In the case of d$_{5/2}$ orbital, the associated quasiparticle
strength is strongly fragmented, and displays 
three low-energy peaks which collect less than 40\% of the single-particle 
strength. In Fig. 3 and  4  we show respectively the energy and the pairing 
gap associated to the lowest of these three peaks.
\end{thebibliography}
\end{document}